\documentclass[nonacm]{acmart}

\AtBeginDocument{%
  \providecommand\BibTeX{{%
    \normalfont B\kern-0.5em{\scshape i\kern-0.25em b}\kern-0.8em\TeX}}}

\usepackage{amsmath,amsfonts}
\usepackage{algorithmic}
\usepackage{tcolorbox}
\usepackage{graphicx}
\usepackage{listings}
\usepackage{makecell}
\usepackage{adjustbox}
\usepackage{multirow}
\usepackage{graphicx}
\usepackage{svg}
\usepackage{xspace}
\usepackage{dsfont}
\usepackage{lipsum}
\usepackage{subfigure}
\usepackage{centernot}
\usepackage{diagbox}
\usepackage{tabularx}

\newcolumntype{P}[1]{>{\centering\arraybackslash}p{#1}}
\usepackage[ruled,vlined,linesnumbered]{algorithm2e}

\SetCommentSty{mycommfont}

\newcommand{\tool}{{\texttt{WireGen}}\xspace}

\usepackage{xcolor}
\usepackage{xcolor,colortbl}
\definecolor{lightgray}{gray}{0.93}
\definecolor{slightgray}{gray}{0.98}
\definecolor{darkgray}{gray}{0.7}

\definecolor{amber}{rgb}{1.0, 0.49, 0.0}

\setcopyright{acmcopyright}
\copyrightyear{2018}
\acmYear{2018}
\acmDOI{XXXXXXX.XXXXXXX}

\acmConference[Conference acronym 'XX]{Make sure to enter the correct
  conference title from your rights confirmation emai}{June 03--05,
  2018}{Woodstock, NY}
\acmPrice{15.00}
\acmISBN{978-1-4503-XXXX-X/18/06}

\begin{document}

\title[Wireframing UI Design Intent with Generative Large Language Models]{Designing with Language: Wireframing UI Design Intent with Generative Large Language Models}

\author{Sidong Feng}
\affiliation{%
  \institution{Monash University}
  \country{Australia}}
\email{Sidong.Feng@monash.edu}

\author{Mingyue Yuan}
\affiliation{%
  \institution{University of New South Wales}
  \country{Australia}}
\email{Mingyue.Yuan@unsw.edu.au}

\author{Jieshan Chen}
\affiliation{%
  \institution{CSIRO's Data61}
  \country{Australia}}
\email{Jieshan.Chen@data61.csiro.au}

\author{Zhenchang Xing}
\affiliation{%
  \institution{CSIRO's Data61 \& Australian National University}
  \country{Australia}}
\email{Zhenchang.Xing@data61.csiro.au}

\author{Chunyang Chen}
\affiliation{%
  \institution{Monash University}
  \country{Australia}}
\email{Chunyang.Chen@monash.edu}

\renewcommand{\shortauthors}{Feng et al.}

\begin{abstract}
Wireframing is a critical step in the UI design process. Mid-fidelity wireframes offer more impactful and engaging visuals compared to low-fidelity versions. However, their creation can be time-consuming and labor-intensive, requiring the addition of actual content and semantic icons. In this paper, we introduce a novel solution \tool, to automatically generate mid-fidelity wireframes with just a brief design intent description using the generative Large Language Models (LLMs). Our experiments demonstrate the effectiveness of \tool in producing 77.5\% significantly better wireframes, outperforming two widely-used in-context learning baselines. A user study with 5 designers further validates its real-world usefulness, highlighting its potential value to enhance UI design process.
\end{abstract}

\ccsdesc[500]{Human-centered computing~Human computer interaction (HCI)}
\keywords{UI wireframe, large language model, fine-tune}

\begin{teaserfigure}
\includegraphics[width=\textwidth]{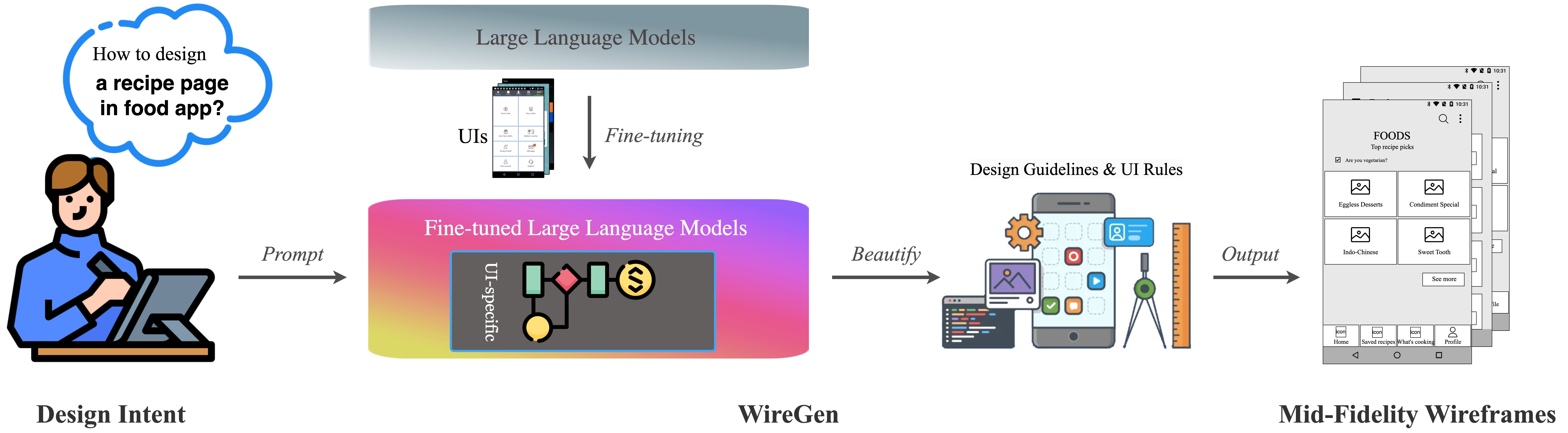}
  \caption{The overview of \tool to streamline UI design process with automated mid-fidelity wireframe generation. We leverage the power of the generative Large Language Models (LLMs), fine-tuning it with thousands of UI data to endow its understanding to UI design knowledge. With this fine-tuned LLMs, designers can simply describe their design intent, and \tool will generate creative mid-fidelity wireframes with ease. }
  \label{fig:teaser}
\end{teaserfigure}

\maketitle

\section{Introduction}
User Interface (UI) plays a crucial role in today's desktop software, mobile applications, and online websites. 
It serves as a visual bridge between a software application and end-users through which they can interact with each other~\cite{feng2022efficiency,feng2023video2action}. 
Good UI designs are essential to the success of a software application and can gain loyalty from the software users~\cite{chen2023unveiling}.
However, designing a good UI can be challenging, even for experienced designers.
On the one hand, designers need to strive for creative ideas and follow many design rules and principles, such as fluent interactivity, universal usability, clear readability, aesthetic appearance, and consistent styles~\cite{galitz2007essential, clifton2015android}.
On the other hand, designers need to rapidly sketch the designs to validate the prototypes, solicit higher-level user feedback, and figure out flaws in the early stage~\cite{chen2020wireframe}.
For this purpose, one common and effective way is wireframing the design intent.

\begin{figure}
	\centering
	\includegraphics[width = 0.98\linewidth]{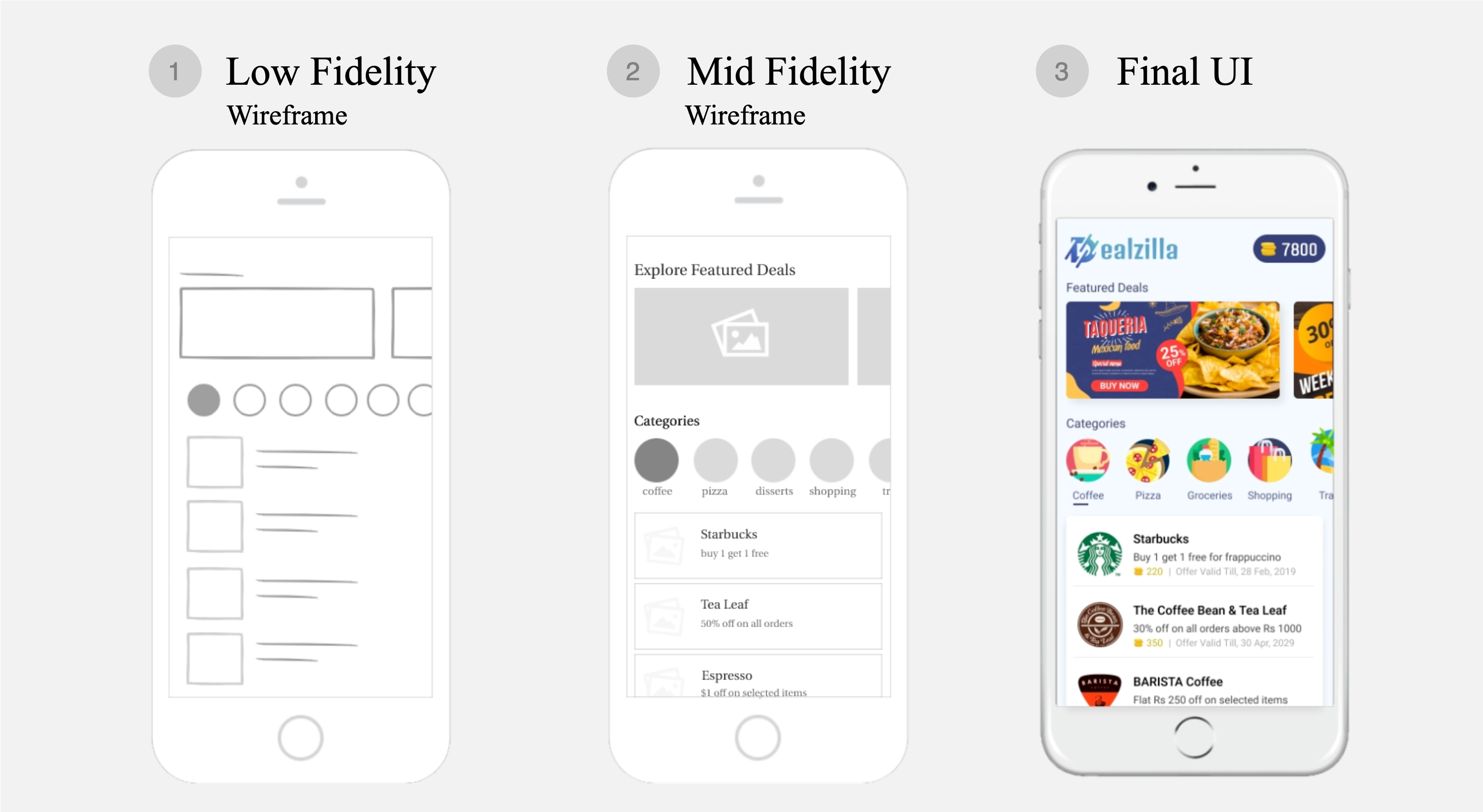}
	\caption{Examples of a low-fidelity wireframe, mid-fidelity UI wireframe, and final UI. }
	\label{fig:wireframe}
\end{figure}

UI wireframe is a basic representation of design intent with a rough sketch, intentionally devoid of colors, graphics, and stylized fonts (see Fig.~\ref{fig:wireframe}). 
This is a quick and cheap way to provide a clear overview of structure, functionality, layout, information flow, and possible user behavior when interacting with the app~\cite{lawson2006designers}.
Typically, designers start from low-fidelity wireframes, the initial visual representation of their ideas with simple layouts and placeholder elements.
Landay~\cite{landay2001sketching} introduces the first interactive tool SILK for low-fidelity UI wireframing.
Huang et al.~\cite{huang2019swire} propose Swire, a system that leverages a deep-learning model to search for similar UI screens from a low-fidelity wireframe, to help designers gain inspiration.
Many studies~\cite{sermuga2022lofi,leiva2020enrico,bunian2021vins,deka2017rico,sermuga2021uisketch,pandian2020syn} have attempted to develop intelligence tools to support low-fidelity wireframes.

However, low-fidelity wireframes can sometimes be too rudimentary to provide an accurate representation of the final product and may need to be modified to account for real-world constraints~\cite{wimmer2020sketchinginterfaces}.
For example, if an original low-fidelity wireframe envisions images and text arranged side by side, but the actual image is too large or the text is too lengthy, it would then need to be altered to a top-to-bottom layout.
To address this, designers refine low-fidelity wireframes into mid-fidelity wireframes by incorporating more details, such as relevant written content, interactive semantic icons, and other elements.
With higher levels of detail, mid-fidelity wireframes allow for more authentic and complex interactions to be explored and can lead to a better final UI, but also require more time and effort to create.
None of the previous works have focused on streamlining the design process for mid-fidelity wireframes.

In this study, we introduce \tool, a novel solution to expedite the UI design process by generating mid-fidelity wireframes from high-level design intent descriptions.
To achieve this, we utilize the stunning generative Large Language Models (LLMs) that inherit billions of web resources, such as web DOMs, semantic relationship, etc.
Since the LLMs are not specifically designed to generate UI wireframes, we fine-tune the LLMs with thousands of UI screens and their corresponding view hierarchies from Rico~\cite{deka2017rico} to help recognize the patterns of the UIs.
With this fine-tuned LLMs, we prompt it in the same way to create innovative wireframes for the test design intent descriptions.
However, UI wireframe design is not a straightforward task, it is subject to design rules, guidelines, and knowledge to match the psychology of human perception, which machines may not be aware of.
To ensure the generated wireframes meet these standards, we further employ post-processing methods to transform raw generations to beautiful UI wireframes, including adding semantic icons, refining text typography, and adhering to UI guidelines.

To evaluate the performance of our LLMs \tool, we first conduct experiments under the specific UI textual descriptions from Screen2Words~\cite{wang2021screen2words}.
The results show that our fine-tuned \tool achieves the best performance (77.5\% significantly better generations) compared with two widely-used large language in-context learning baselines.
Our \tool also generates on average 84.5\% significantly better UI wireframes, which outperforms three ablation models.
As there are many ways to describe a design intent in different words, we further conduct a user study with five professional designers to gain insight into the usefulness of  our \tool.
The study reveals that designers respond positively to the mid-fidelity wireframes generated by our tool and provide valuable feedback.
Lastly, we discuss the limitations and the implications of \tool. 
Altogether, our paper makes the following contributions:
\begin{itemize}
    \item To the best of our knowledge, this is the first study that investigates the collaboration between humans and AI for creating mid-fidelity wireframes.
    \item We present \tool, that harnesses the power of Large Language Models (LLMs) to generate mid-fidelity wireframes from a simple description of the design intention.
    \item The experiments and user study demonstrate the effectiveness and usefulness of \tool in aiding designers with UI design.
\end{itemize}

\section{Related Work}
\label{sec:related}
We aim to facilitate the design of mobile mid-fidelity UI wireframes by simply prompting the design descriptions using Large Language Models (LLMs). 
To this end, we review the related work in three main areas: 1) mobile UI wireframe, 2) designing with natural language, and 3) prompting Large Language Models.

\subsection{Mobile UI Wireframe}
Wireframing plays a pivotal role in many types of creative activities because of its highly visual nature and its flexibility for creation and manipulation: designers can create and imagine any kind of visual content, and continuously revise it without being constrained by unnecessary details~\cite{goldschmidt1991dialectics,lawson2006designers,xie2022psychologically}.
Subsequently, mobile UI wireframing is well-suited for the early stage of the design process to expand novel ideas, visualize abstract concepts, improve user experience, and rapidly compare alternatives~\cite{lin2000denim, wu2021exploring}.
SILK~\cite{landay2001sketching} is the first system that allows designers to create interactive UI prototypes by wireframing.
With the essential of UI wireframe as a profession in the UI design process, many academic research~\cite{lu2022bridging,huang2021sketch} and commercial graphic tools including Adobe PhotoShop~\cite{web:photoshop}, Sketch~\cite{web:sketch}, and Figma~\cite{web:figma} are developed to allow designers to sketch the UI wireframes at multiple detail levels from low-fidelity to mid-fidelity and final UI (as shown in Fig.~\ref{fig:wireframe}). 
Low-fidelity wireframes serve as the design’s starting point that tends to be fairly rough, created without any sense of scale, grid, or pixel-accuracy to represent a basic visual of the UI.
Mid-fidelity wireframes further boast pixel-specific layouts and make the UIs realistic by including semantic icons and relevant written content, that allows for exploring and documenting complex concepts such as menu systems or interactive buttons.

Recent advances in AI have explored human-AI co-creation research for UI wireframes~\cite{subramonyam2021protoai}.
In these systems, AI serves as humans’ collaborators, that make recommendations based on designers’ goals and intentions.
For example, Swire~\cite{huang2019swire} proposes a deep-learning model to input a low-fidelity UI wireframe and retrieve relevant UI examples from large-scale datasets to help designers gain inspiration effectively.
Lately, researchers conduct data-driven research to support more advanced design inspiration search~\cite{sermuga2022lofi,leiva2020enrico,bunian2021vins,deka2017rico,sermuga2021uisketch,pandian2020syn,feng2022gallery} or layout generation~\cite{li2019layoutgan,gupta2021layouttransformer,arroyo2021variational} in complement to designers’ agency and creativity.
These works, however, still require the designers to expend time and effort designing and sketching a possible low-fidelity UI wireframe, then adding adequately relevant content to demonstrate mid-fidelity UI wireframes.
Our work aims to simplify the process further: given a design intent description, we automatically generate mid-fidelity UI wireframes for designers to broaden their horizons and get inspiration.

\subsection{Designing with Natural Language}
To identify business requirements and collaborate ideas for UI designs, instigating conversations with the stakeholders and users is crucial.
As a subsequent effort, many studies are devoted to bridging natural language and UI screens.
For example, Screen2Vec~\cite{li2021screen2vec} uses a self-supervised approach to learn the representation of a UI screen from multi-modal information such as the textual content, visual design, hierarchy layout patterns, and app meta-data.
Similarly, Screen2Words~\cite{wang2021screen2words} proposes a screen summarization approach to describe the functionalities of the UI screenshots.
Some research specifically generates semantically alt-text labels for UI elements~\cite{feng2023read,feng2022auto,chen2022towards,feng2021auto,feng2022gifdroid,feng2021auto,chen2020lost,xie2020uied,chen2019gallery}.
These studies attempt to learn the latent representation of UIs to generate natural language understanding.
On the contrary, our work focuses on interpreting natural language descriptions into UIs.

The interest in involving natural language as a form of interaction for graphic designs has recently found success in text-to-image generative models.
Variational autoencoder (VAE)~\cite{kingma2013auto} and generative adversarial network (GAN)~\cite{goodfellow2020generative} are frequently used in generating graphic designs.
For example, Aoki et al.~\cite{aoki2022emoballoon} introduce a GAN-based model EmoBalloon to generate emotional speech balloons in the chat UI.
Recently, advances in LLMs ~\cite{brown2020language} and diffusion models~\cite{ho2020denoising} have introduced methods that are remarkable at generating images based on text prompts.
While state-of-the-art text-to-image works such as DALLE~\cite{ramesh2022hierarchical}, Stable Diffusion~\cite{rombach2022high}, and MidJourney~\cite{web:midj} show amazing performance in creating aesthetic designs, they still have risks in generating realistic designs~\cite{web:limitdalle,ramesh2022hierarchical}.
First, they are great at drawing but horrible at spelling words, e.g., prompting with ``an image with Twitter text'' may generate ``Ttiter'', ``Tw.uTe:'', or mostly unrecognizable text.
Second, coherence in designs is often missing while human creations would never lack, e.g., shape incoherences, lack of components composability, etc.
These potential flaws make it difficult to apply to create UI wireframes with the finest details as demonstrated in Fig.~\ref{fig:comparison}.
As a substitution, Huang et al.~\cite{huang2021creating} propose a UI Generator, that uses a deep-learning model to generate coordinates of UI elements from textual description.
However, those coordinates can only be interpreted into the layout of UI, i.e., a simple low-fidelity UI wireframe, which still requires designers to mentally imagine the actual contents in their heads.
Our work expands their research by prompting generative LLMs, comprising billions of web layout and content understandings, to generate a UI-specific language, which can be parsed into a mid-fidelity wireframe, including similar-to-real content, semantic functionalities, etc.

\subsection{Prompting Large Language Models}
Deep learning has introduced more opportunities for natural language understanding.
The development of transformer-based deep neural language models such as BERT~\cite{devlin2018bert} has shown its potential value in many applications.
For example, the aforementioned UI Generator~\cite{huang2021creating} leverages a BERT model to embed text description to generate a sequence of UI elements' coordinates.
Recent advancements in large language models (LLMs), such as GPT~\cite{brown2020language}, LLaMA~\cite{touvron2023llama}, PaLM~\cite{chowdhery2022palm}, RoBERTa~\cite{liu2019roberta}, have led to boost performances on zero-shot, few-shot, and fine-tuned with handcrafted prompts compared to prior deep learning methods.
Zero-shot prompts directly describe what ought to happen in a task, and few-shot prompts show the LLMs what pattern to follow by feeding it examples of desired inputs and outputs.
While zero-shot or few-shot prompts allow to prototype common tasks, their inherent limitations (e.g., lack of domain-specific understanding, limited input prompt length) make them less capable of prototyping specific applications~\cite{liu2021gpt,moradi2021gpt,yang2022empirical}.
Therefore, LLMs fine-tuning, encoding lexical, syntactic, and semantic regularities of the domain-specific language, is then used to master specific task capabilities.
For example, Codex~\cite{chen2021evaluating} is a fine-tuned version of GPT-3, that can help developers with code generation.
Many studies have applied fine-tuning to support different domain-specific tasks, such as InstructGPT~\cite{ouyang2022training}, FLAN-T5~\cite{chung2022scaling}, math word solving~\cite{zong2022solving}, etc.
In this same line of research, we fine-tune the LLMs to support the domain-specific task of UI wireframing, providing insight into how natural language interaction can help designers.

\begin{figure*}
	\centering
	\includegraphics[width = 0.98\textwidth]{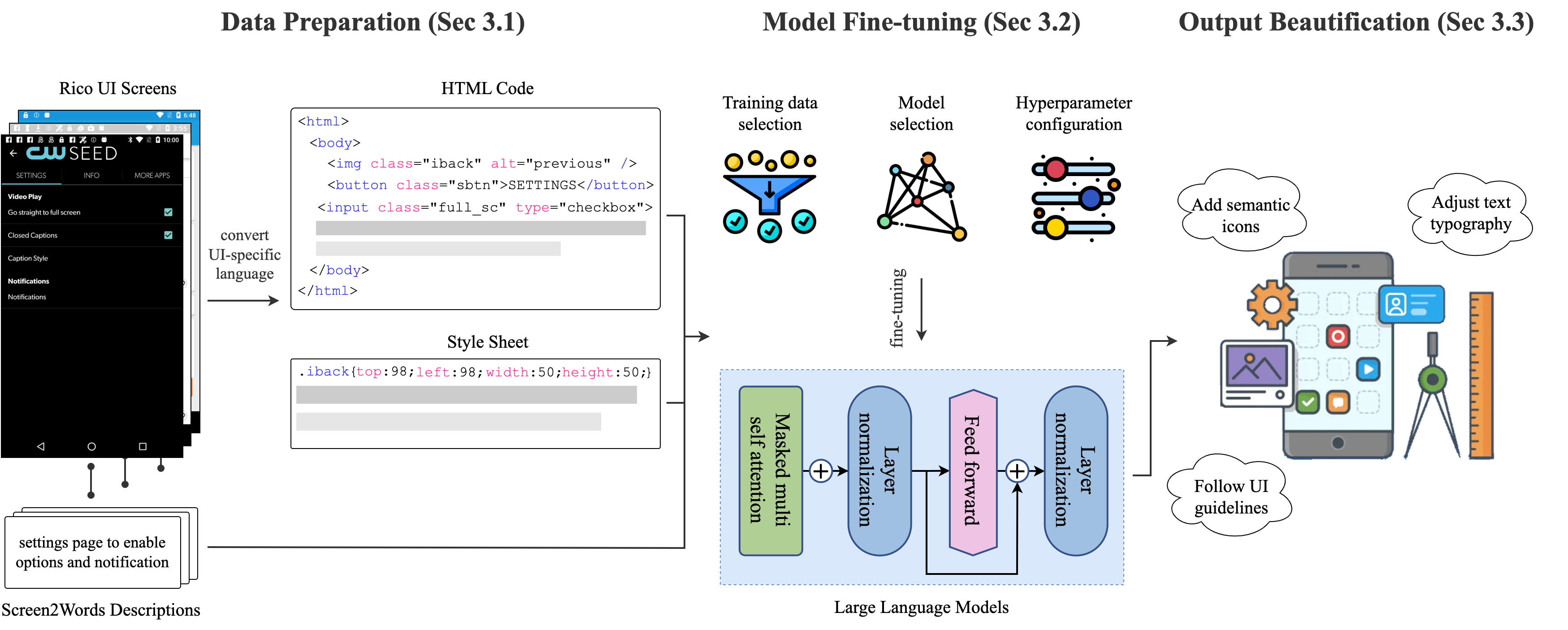}
	\caption{The overview of \tool that contains three phases: (i) \textit{Data Preparation} phase (Section~\ref{sec:dataset}) that collects dataset of UI wireframes and their associated high-level descriptions. (ii) \textit{Model Fine-tuning} phase (Section~\ref{sec:finetune}) that implements the the best practices of modern strategies to fine-tune the LLMs. (iii) \textit{Output Beautification} phase (Section~\ref{sec:post}) that transforms the raw output into visually appealing UI wireframes.}
	\label{fig:overview}
\end{figure*}

\section{Approach}
Given a natural language description of the design intent for a UI, we harness the power of generative LLMs to automatically create mid-fidelity wireframes to greatly help designers gain design inspiration and create prototypes for different use cases and contexts.
The overview of our approach is shown in Fig.~\ref{fig:overview}, which is divided into three main phases: 
(i) \textit{Data Preparation} phase that collects dataset of UI wireframes and their associated high-level descriptions;
(ii) \textit{Model Fine-tuning} phase that implements the the best practices of modern strategies to fine-tune the LLMs to recognize the UI patterns;
(iii) \textit{Output Beautification} phase that transforms the raw output into visually appealing UI wireframes that are interactive and consistent with design guidelines.

\subsection{Prepare Dataset of UI Wireframes and Descriptions}
\label{sec:dataset}
\subsubsection{Mid-fidelity Wireframes and Descriptions}
While there are many low-fidelity wireframe datasets available~\cite{sermuga2022lofi,leiva2020enrico,bunian2021vins,deka2017rico,sermuga2021uisketch}, none of the prior work studies on mid-fidelity wireframes.
This is because mid-fidelity wireframes go beyond the shape, placeholders, and ``lorem ipsum'' text of low-fidelity wireframes to include real content, semantic icons, interactive elements, etc.
However, manual labeling of wireframes can be prohibitively expensive.
To that end, we propose a novel approach to automatically collect a set of representations of mid-fidelity wireframes.
As shown in Fig.~\ref{fig:wireframe}, mid-fidelity wireframes are typically a simplified and monochromatic representation of the final UI.
With this in mind, we propose to gather representations of mid-fidelity wireframes from existing UI screen datasets.

We utilize one of the largest open-sourced UI datasets, Rico~\cite{deka2017rico}.
The Rico dataset contains 66k unique UI screens from more than 9.7k Android apps across 27 diverse app categories.
The dataset provides a screenshot image and a view hierarchy of UI objects.
Each object has a set of properties, including its resource id, type (e.g., Button, Image, Text, etc.), bounding box location on the screen, textual content (if any), and various other properties such as clickability, scrollability, etc.
These screens can serve as our mid-fidelity wireframes.

To gather the descriptions of these screens, we utilize a comprehensive screen summarization dataset Screen2Words~\cite{wang2021screen2words}, that captures the complex information of UIs into concise language descriptions.
The summarization is done by 85 professional labelers through a rigorous labeling process and guidelines, resulting in high consistency in both linguistic coherence and on-screen focus area among the labelers.
As a result, we construct a dataset of pairs of UI screens and their corresponding textual descriptions.

\subsubsection{UI-specific Language}
One challenge with using LLMs is that they can only process text input\footnote{Since OpenAI did not release its multimodal API (GPT-4) before submission, it was difficult for us to measure the capability of LLMs for UI screen understanding of complex structures, semantic icons, etc.}, while UI screens are multimodal, containing text, semantic icons, structural information, etc.
To help LLMs inherently understand UI screens, we aim to convert the UI screens into domain-specific language that LLMs can understand, which is known as prompt engineering~\cite{reynolds2021prompt}.
A well-designed prompt helps the LLMs elicit specific knowledge and abstractions needed to complete the task.
Since the training samples of LLMs are typically scraped from the raw web page data, e.g., GPT was trained on 410 billion tokens from the Common Crawl web corpus, we use HTML syntax as the domain-specific language to convert the UI screens into text.
The closer the prompt is semantically similar to the LLMs' training samples, the better the inference will be.

To translate the UI screen with view hierarchy into HTML syntax, we need to preserve the properties and structural relationship of UI elements.
While the view hierarchy resembles a DOM tree in HTML, e.g., it starts with a root view and contains UI elements descending in the tree, there are two fundamental limitations.
First, the native classes in the view hierarchy don't always match HTML tags, for instance, a \textit{<RadioButton>} in the hierarchy corresponds to a combination of \textit{<input type="radio">} and \textit{<label>}.
Second, including all properties of UI elements will result in excessively long HTML text that may exceed the maximum input token length of LLMs.

\begin{figure*}
	\centering
	\includegraphics[width = 0.9\textwidth]{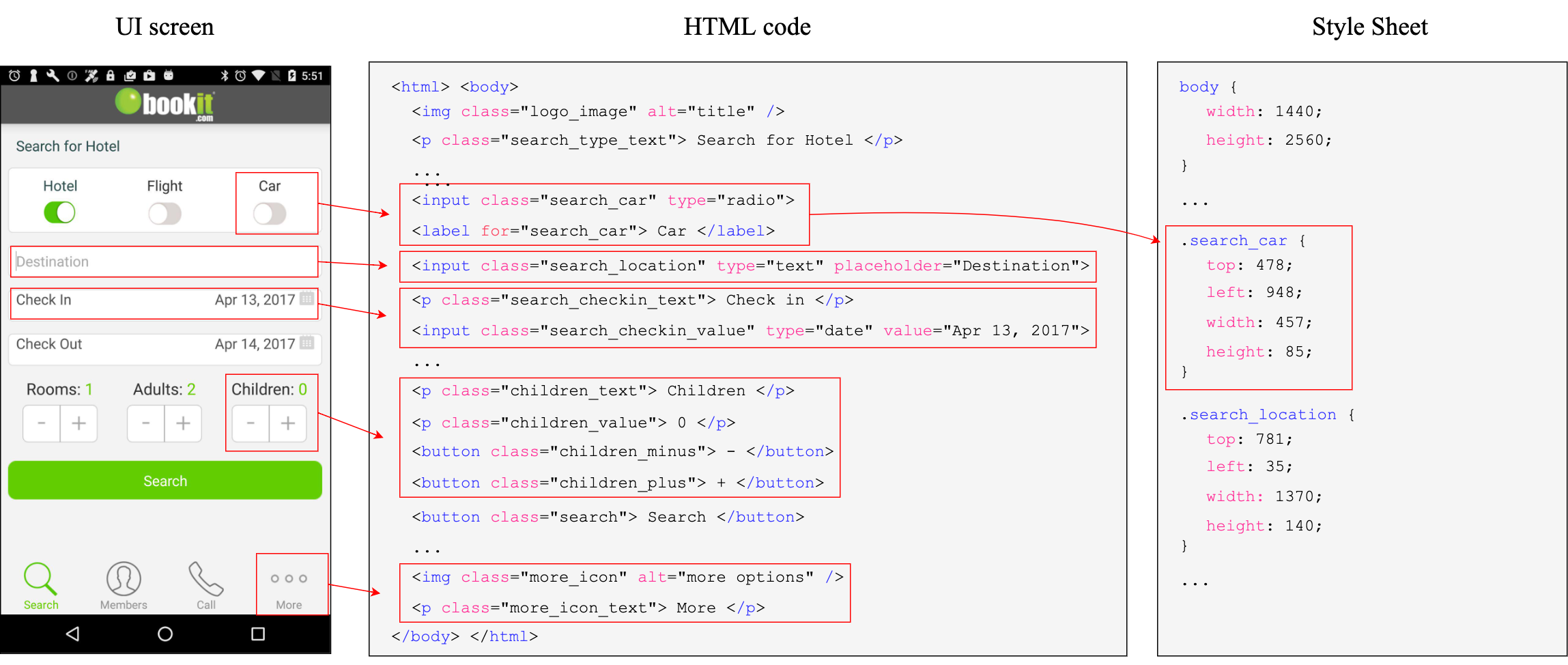}
	\caption{Example of converting the UI screen to HTML code and style sheet.}
	\label{fig:prompt}
\end{figure*}

\renewcommand{\arraystretch}{1.25}
\begin{table}
	\centering
        \footnotesize
        \tabcolsep=0.18cm
	\caption{The class conversion between view hierarchy and HTML syntax.}
	\label{tab:conversion}
	\begin{tabularx}{0.73\linewidth}{p{2.15cm}|p{2.75cm}|X} 
		\hline
		\rowcolor{darkgray} \bf{CLASS} & \bf{FUNCTIONALITY}  & \bf{HTML CODE} \\
		\hline
		\hline
		\rowcolor{lightgray} TextView & display text content & \textit{<p class=$id$> $text$ </p>} \\
		  \hline
		  \rowcolor{slightgray} \makecell[Xt]{Button, \\ ToggleButton} & click to new events & \textit{<button class=$id$> $text$ </button>} \\
		  \hline
		  \rowcolor{lightgray} \makecell[Xt]{ImageView, \\ ImageButton} & display image & \textit{<img class=$id$ alt=$content$ />} \\
		\hline
		  \rowcolor{slightgray} EditText & input text & \textit{<input class=$id$ placeholder=$text$ type=``text''>} \\
		\hline
            \rowcolor{lightgray} CheckBox, Switch & (de)select an option & \textit{\makecell[Xt]{<input class=$id$ type=``checkbox''> \\ <label for=$id$> $text$ </label>}}  \\
            \hline
            \rowcolor{slightgray} RadioButton & choose only one option & \textit{\makecell[Xt]{<input class=$id$ type=``radio''> \\ <label for=$id$> $text$ </label>}} \\
		  \hline
            \rowcolor{lightgray} DatePicker & select a date & \textit{<input class=$id$ type=``date'' value=$text$>} \\
            \hline
            \rowcolor{slightgray} Spinner & select an option from drop-down menu & \textit{\makecell[Xt]{<select class=$id$ type=``radio''> </select> \\ <label for=$id$> $text$ </label>}} \\
            \hline
            \rowcolor{lightgray} VideoView & display video & \textit{<video class=$id$ alt=$content$> </video>} \\
            \hline
            \rowcolor{slightgray} Other & less commonly-used classes & \textit{<div class=$id$> </div>} \\
            \hline
	\end{tabularx}
\end{table}

Therefore, we adopt a similar approach to previous work~\cite{feng2023prompting,wang2022enabling} to convert the view hierarchy of the UI screen into HTML syntax with similar functionality.
An example is shown in Fig.~\ref{fig:prompt}. 
In detail, we first adopt a depth-first search traversal algorithm to iterate through each node starting from the root of the view hierarchy.
During the iteration, we convert each node into HTML code and style sheet, based on a selected subset of properties from the view hierarchy.
\begin{itemize}
    \item \texttt{resource\_id}: describes the unique resource id of the element, depicting the referenced resource.
    \item \texttt{class}: describes the native UI element type such as TextView and Button.
    \item \texttt{text}: describes the text of the element (if any).
    \item \texttt{content\_desc}: conveys the content descriptions of the visual element such as ImageView and VideoView.
    \item \texttt{bounds}: describes the positional information of the element such as top, left, width, and height.
\end{itemize}

We develop a heuristic approach based on Table~\ref{tab:conversion} to match the classes in the view hierarchy to HTML tags with equivalent functions.
For instance, the \textit{<TextView>} is mapped to the \textit{<p>} tag; buttons are mapped to the \textit{<button>} tag; and image elements to the \textit{<img>} tag.
Unlike the classes in the view hierarchy, HTML uses the combination of \textit{<input>} and \textit{<label>} for input-related elements (e.g., \textit{<EditText>}, \textit{<CheckBox>}, \textit{<RadioButton>}, etc.).
The \textit{<input>} represents the specific class and \textit{<label>} represents the text property.
Note that we focus on the most commonly-used classes for simplicity, and the rest of the classes, including containers such as \textit{<LinearLayout>}, are mapped to the \textit{<div>} tag.

Next, we insert properties into the HTML code following standard syntax.
For instance, the unique identifier \textit{class} for the objects in the HTML code is set using the resource\_id.
\texttt{Text} properties are placed between the opening and closing HTML tags.
Since the \textit{<EditText>} usually depicts text in the \textit{placeholder}, we replace it accordingly in Table~\ref{tab:conversion}.
For image-related objects, the \textit{alt}-text is conveyed using the content\_desc property.

To describe the style and layout of the UI, we generate a style sheet in addition to the HTML code.
To precisely generate the layout, we use the \texttt{bounds} property in the view hierarchy to encode the absolute position of each atomic element, including the \textit{top}, \textit{left}, \textit{width}, and \textit{height}.
An example of style sheet is shown in Fig.~\ref{fig:prompt}.
We avoid using relative positioning, such as inline or margin, as it could limit the scope of adjacent elements~\cite{web:css}.
Note that we also add the overall width and height of the UI screen in the style sheet.

\subsection{Fine-tune Large Language Models with Best Practice}
\label{sec:finetune}
Since the LLMs are not specifically designed to understand the UI design patterns, we fine-tune the LLMs with the natural language description as the input and the UI wireframe in HTML syntax as the output (the same learning objective as the pre-trained model).
The implementation of fine-tuning process is not a trivial task~\cite{lecun2015deep}, which can significantly influence the performance of LLMs.
To that end, we summarize three highly sensitive aspects in fine-tuning LLMs, denoting training data selection, model selection, and hyperparameter configuration, and detail our implementation informed by the best practices.

\subsubsection{Training data selection}
When fine-tuning LLMs, the size of the training dataset plays a non-negligible role~\cite{mehrafarin2022importance}.
On the one hand, the data size needs to be sufficient, diverse, and representative to enable the LLMs to learn the characteristics of the specific task effectively.
On the other hand, training on an excessive dataset can be time-consuming and costly\footnote{The pricing set by OpenAI \url{https://openai.com/api/pricing/}}.
To strike a balance between the LLMs' capability and the training cost, we attempt to select 1,000 samples from the dataset, suggested by the previous work~\cite{zong2022survey}.

Regarding the sample selection, a simple random selection cannot ensure the LLMs' generalizability and diversity, as the UIs in the same app may convert to very similar HTML syntax. 
To avoid this data leakage problem~\cite{kaufman2012leakage}, we select the screens in the dataset by the app. 
We also ensure the representation of app categories and screen summaries are diverse in the training dataset.
In total, we collect 1,000 samples from 191 apps, covering 27 app categories, summarised by an average of 7.1 words.

\subsubsection{Model selection}
There are numerous emerging LLMs that have exhibit promising performance in natural language understanding and logical reasoning, such as PaLM~\cite{chowdhery2022palm}, RoBERTa~\cite{liu2019roberta}, T5~\cite{raffel2020exploring}, etc.
In this work, we adopt the recent state-of-the-art LLM, GPT~\cite{brown2020language} from OpenAI with 175 billion parameters pre-trained on a massive dataset.
It is based on the transformer model~\cite{vaswani2017attention} including masked multi-self attention, normalization layers, and feed-forward layers (see in Fig.~\ref{fig:overview}).
GPT offers sets of models to support different levels of tasks, including Curie, Babbage, Ada, etc.
For our study, we choose the model Turbo (gpt-3.5-turbo) which is well suited to our task due to two reasons.
First, Turbo is the most advanced model and can perform tasks with less instruction compared to other models.
It is especially ideal for tasks of creative content generation and extensive understanding of the content.
Second, Turbo excels in solving logic problems and understanding the intent of code (e.g., HTML), and has been fine-tuned for programming applications like Codex.

\subsubsection{Hyperparameter configuration}
In addition to the choice of model, fine-tuning performance can also be improved through the customization of hyperparameters.
According to the previous fine-tuning practices of LLMs~\cite{brown2020language,web:practice} that shows promising performance across a range of use cases, we apply a learning rate of 0.1 and a batch size of 256.
We train the model for 4 epochs to enable the model to recognize the input prompt syntactic and usage patterns of the fine-tuning data.
Furthermore, there are three additional hyperparameters that have been optimized for the LLMs, specifically GPT model:

\begin{itemize}
    \item \texttt{Temperature:} determines how much randomness is in the generation, regarding new content creation. At a \texttt{temperature} of 0.0, the model will always produce the same fixed response to an input text, regardless of the number of generations.  Raising the \texttt{temperature} value (with a maximum of 1.0) enables more creative output from the pre-trained resources, but also increases the risk of generating irrelevant output. Based on previous studies~\cite{brown2020language, zong2022survey}, we set the value to 0.65 to balance robustness and creativity.
    \item \texttt{Maximum\_length:} sets the upper limit of the number of generated tokens. The default value is 256, but generating HTML syntax often requires more. Therefore, we set the value to the maximum of 4,096 tokens to accommodate the requirement.
    \item \texttt{Stop\_sequence:} when the generation of tokens stops. To control the endpoint of the generation in our task of HTML syntax generation, As we aim to generate HTML syntax, we set the delimiter value to the HTML closing tag, \textit{</html>}.
\end{itemize}

\subsection{Transform Raw Generation into Beautiful UI Wireframe}
\label{sec:post}
\begin{figure}
	\centering
	\includegraphics[width = 0.60\linewidth]{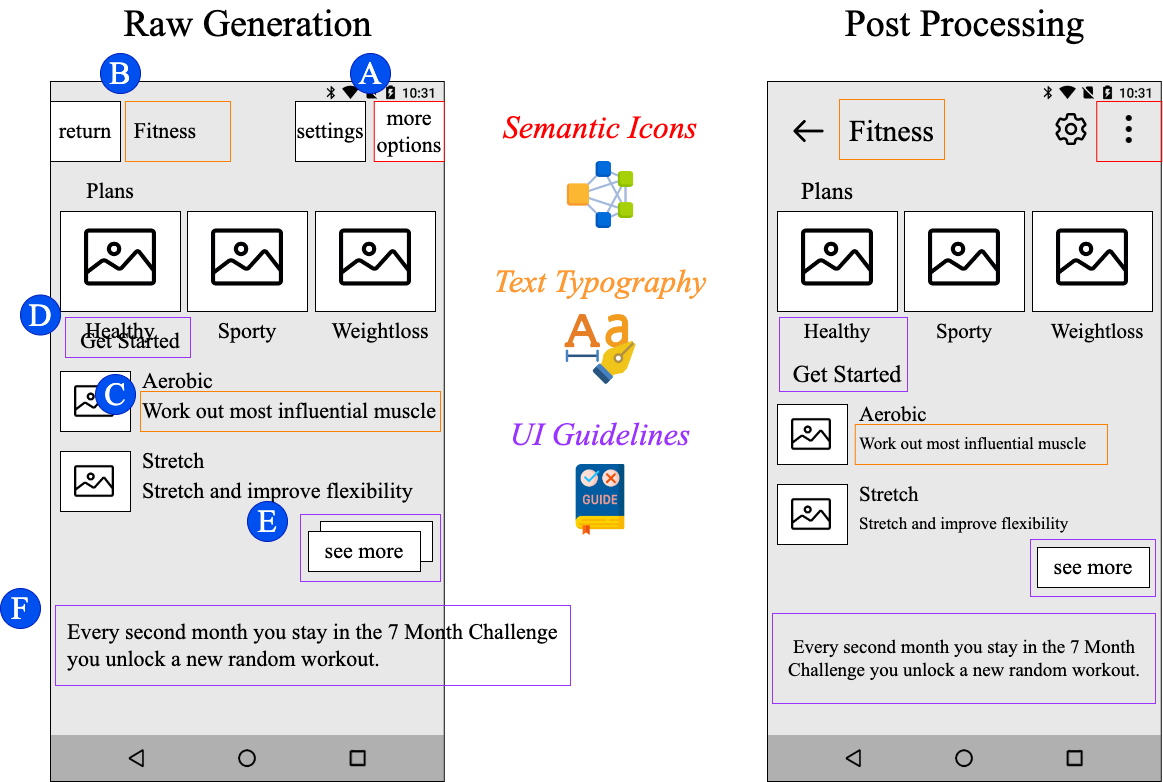}
	\caption{An illustration of post-processing raw generation into better UI wirframes, adding semantic icons, refining text typography, and adhering to UI guidelines.}
	\label{fig:post}
\end{figure}

After fine-tuning process, we prompt the LLMs with a simple description of design intent to generate a HTML syntax of mid-fidelity wireframes.
However, programming languages like HTML and CSS are not easy for designers to understand.
We wish to hide the programming hardships under-the-hood to reduce the designer’s burden.
Besides, the LLMs may not fully capture the intricacies of the design knowledge behind the UI designs.
To this goal, we propose post-processing methods to transform the raw HTML syntax generation into intuitive, fluent, and interactive UI wireframes.
An example of the post-processing methods is shown in Fig.~\ref{fig:post}.

\begin{table*}
\caption{The 10 most common icon semantics identified through an iterative open coding of Rico.}
\centering
\small
\begin{tabular}{p{0.12\textwidth}p{0.45\textwidth}p{0.3\textwidth}}
    \toprule
    \bf{ICON} & \bf{ASSOCIATED SEMANTICS} & \bf{EXAMPLES} \\
    \midrule
    \parbox[c]{1em}{\includegraphics[width=0.13in]{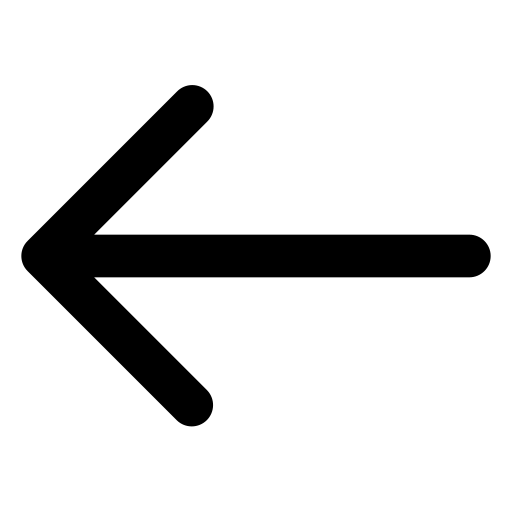}} & \small{return, back, navigate up, previous, backwards, arrow back} & \parbox[c]{1em}{\includegraphics[width=1.8in]{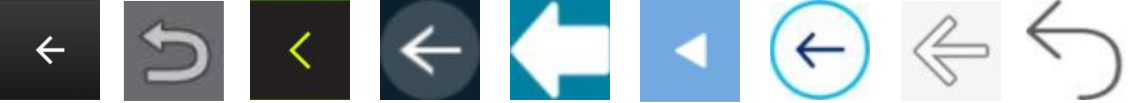}} \\

    \parbox[c]{1em}{\includegraphics[width=0.13in]{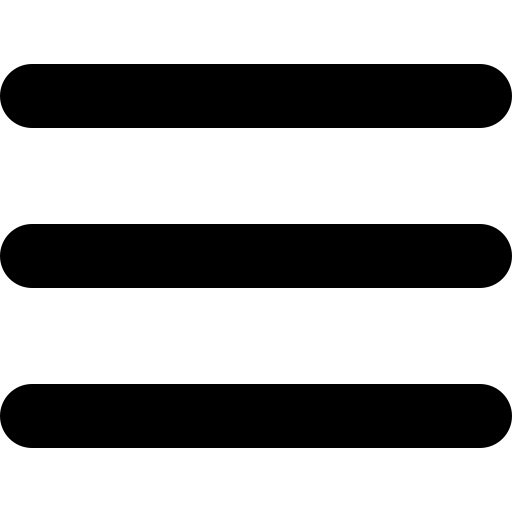}} & \small{menu, navigation drawer, list, card, dashboard} & \parbox[c]{1em}{\includegraphics[width=1.8in]{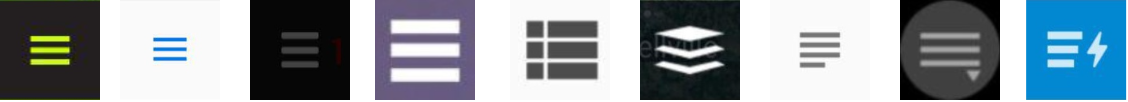}} \\

    \parbox[c]{1em}{\includegraphics[width=0.13in]{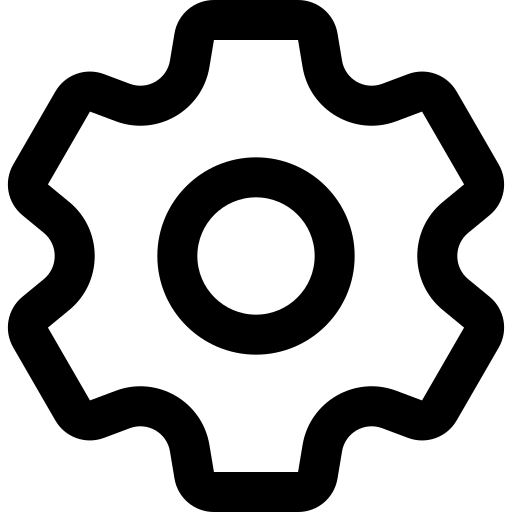}} & \small{settings, toolbox, gear, preferences, options} & \parbox[c]{1em}{\includegraphics[width=1.8in]{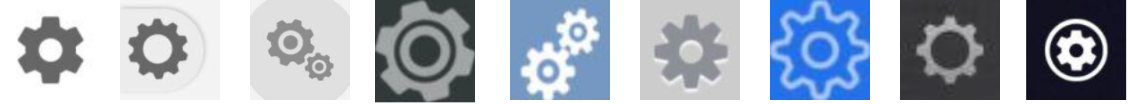}} \\

    \parbox[c]{1em}{\includegraphics[width=0.13in]{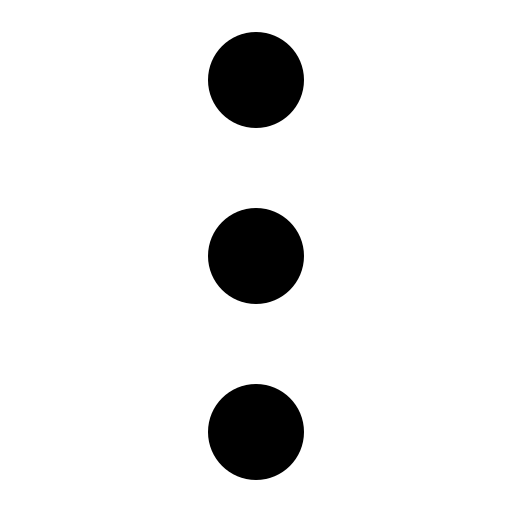}} & \small{more, more options, dots, three, overflow} & \parbox[c]{1em}{\includegraphics[width=1.8in]{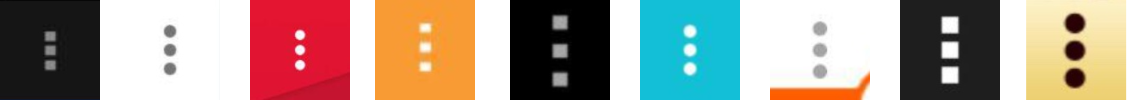}} \\
    
    \parbox[c]{1em}{\includegraphics[width=0.13in]{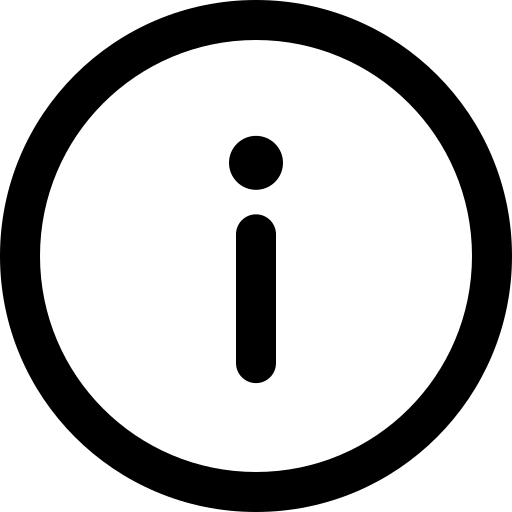}} & \small{information, info, help, support, question, ask, faq } & \parbox[c]{1em}{\includegraphics[width=1.8in]{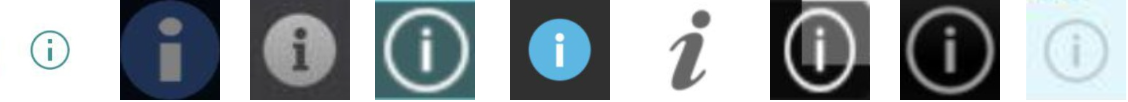}} \\

    \parbox[c]{1em}{\includegraphics[width=0.13in]{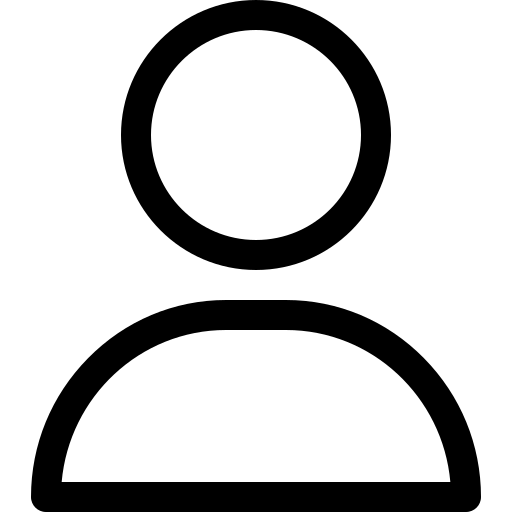}} & \small{person, user, avatar, account, customer, profile} & \parbox[c]{1em}{\includegraphics[width=1.8in]{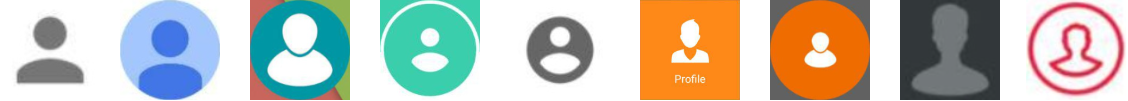}} \\

    \parbox[c]{1em}{\includegraphics[width=0.13in]{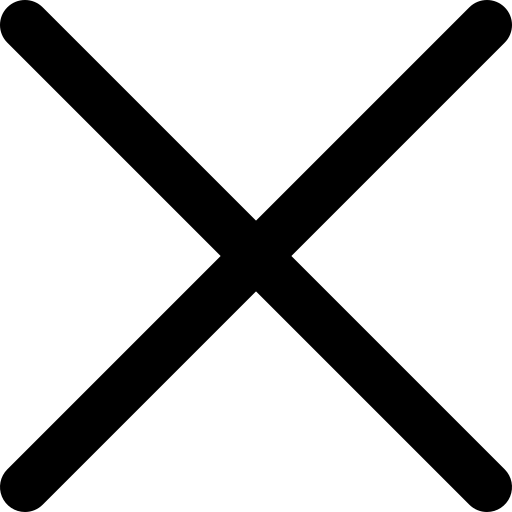}} & \small{close, quit, logout, exit, switch-off} & \parbox[c]{1em}{\includegraphics[width=1.8in]{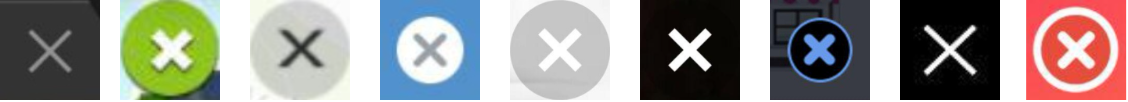}} \\

    \parbox[c]{1em}{\includegraphics[width=0.13in]{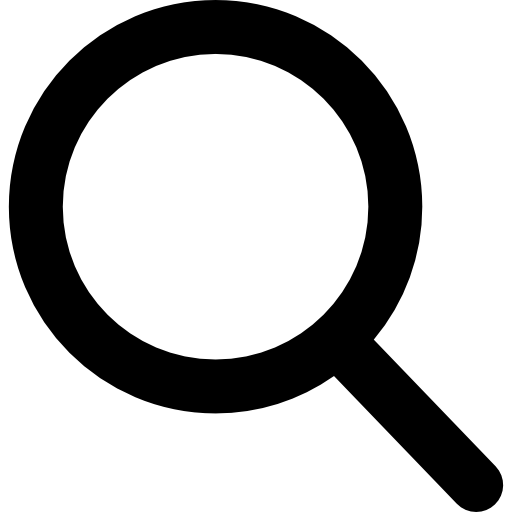}} & \small{search, investigate, search-engine, magnifier, find, glass } & \parbox[c]{1em}{\includegraphics[width=1.8in]{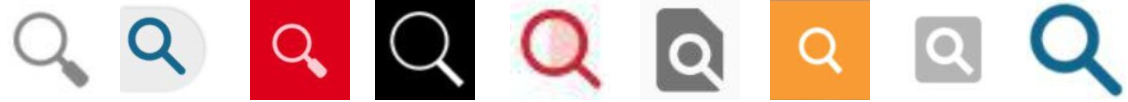}} \\

    \parbox[c]{1em}{\includegraphics[width=0.13in]{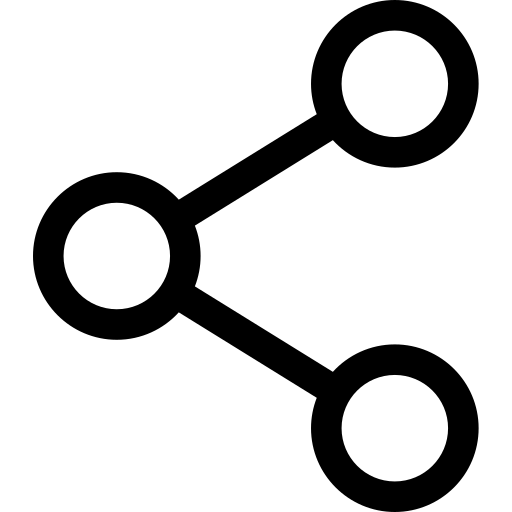}} & \small{share, share button, forward, social media} & \parbox[c]{1em}{\includegraphics[width=1.8in]{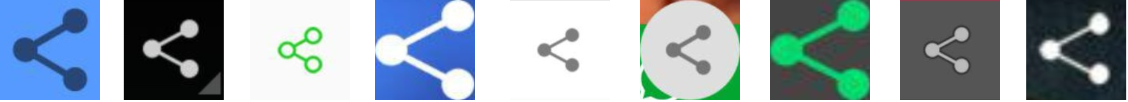}} \\

    \parbox[c]{1em}{\includegraphics[width=0.13in]{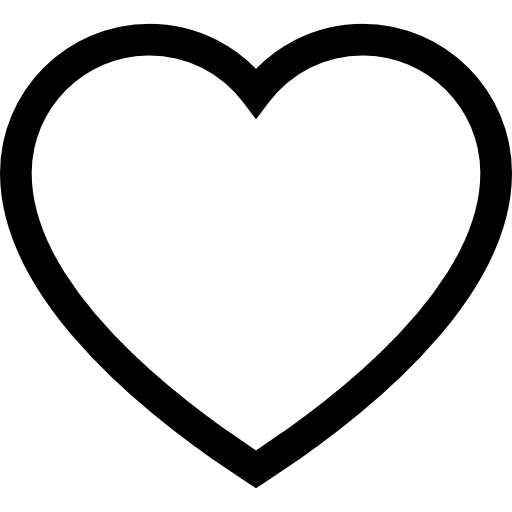}} & \small{favourite, like, heart, upvote} & \parbox[c]{1em}{\includegraphics[width=1.8in]{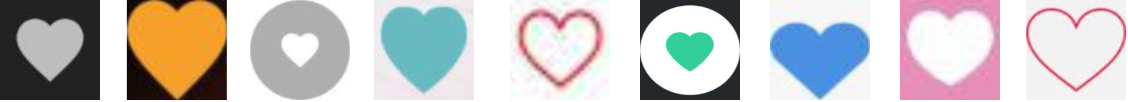}} \\

    \bottomrule
\end{tabular}
\label{tab:categorization}
\end{table*}

\subsubsection{Semantic Icons}
Different from the buttons with text that explicitly shows the functionalities, the icons fail to expose the textual semantics.
To address this, we add the semantic icons in the wireframe by leveraging the \textit{alt} text in the HTML syntax.
First, we carry out a small study to understand the icons and their associated text semantics.
Based on the Rico dataset, we extract a large number (73,449) of icons and randomly select 4,000 (5\%) as our experimental set.
To identify the set of frequently occurred icon semantics in the wireframe, we perform an iterative open coding of the experimental set using the existing expert lexicon of categories in books and websites such as Google’s Material icon set~\cite{googlematerial} and IBM’s Design Language of Iconography~\cite{ibmicon}.
Two researchers from our team independently code the categories of these icons, recording any part of the initial vocabulary.
Note that both researchers have design experience in UI wireframe design.
After the initial coding, the researchers meet and discuss the discrepancies and the set of new semantic categories until a consensus is reached.
A category of top 10 icon semantics can be seen in Table~\ref{tab:categorization}.
For example, the $\leftarrow$ icon usually expresses the semantic of ``return'' and the $\times$ icon usually expresses the semantic of ``close''.

Based on the icon semantic category, we prompt the LLMs to convert the alt-text description in the HTML syntax into the corresponding icons.
In detail, we first provide the LLMs with the context of our icon semantic category with the prompt, i.e., \textit{``Here is an icon semantic category: first icon can be assigned an alternative description of return, back, navigate up, previous, ...''}.
Next, we use a prompt to instruct the LLMs to identify each icon image with its \textit{alt}-text attribute in the HTML syntax.
For instance, in Fig.~\ref{fig:post}-A, to generate the icon for the alt-text of ``more options'' at the top right corner, we prompt the LLMs by \textit{``Please indicate the icon number if there is a corresponding icon for the alternative description of ``more options''. If there is no related icon, please respond with no.''}
In this case, the LLMs returns the ``fourth'' icon in the category, which corresponds to the ``three-dots'' icon as shown in the Fig.~\ref{fig:post}-A.
Note that we do not perform LLMs fine-tuning for this task, as the icon identification is relatively straightforward, either select a single icon from the category or indicate no relevant icons.

\subsubsection{Text Typography}
The typography of text layout is defined as the process of overlaying texts onto the text blocks in the UI wireframe.
However, the process poses potential challenges~\cite{yang2016automatic,alves2022towards} on text wrapping, text alignment, font size, etc.
We define text wrapping $S = \{s_1, s_2, ..., s_{n}\}$ consists of $2^{|s|-1}$ possible wrapping ways on string $s$; text alignment $A = \{$\textit{``left'', ``center'', ``right''}$\}$ consists of the alignment ways of text in block; $F = \{$\textit{``small'', ``medium'', ``large''}$\}$ to demonstrate the importance of the text and each font has a number of corresponding sizes (i.e., \textit{``normal''} $\in \{10,11,...\}$); $w$ and $h$ denotes the width and height of the text block.
To ensure the text typography aligns with human design principles, we formulate it as an optimization problem that minimizes the waste of spare block space and the mismatch of information importance in perception and semantics.
\begin{equation}
	optimize\textit{(s | w, h)} = max(\frac{S_i * F_j}{w*h}) 
	\label{eq:sim}
\end{equation}
In detail, we first identify the text font  by the \textit{class} in the HTML syntax, that $F$ corresponds to the title, normal text, and subtitle for \textit{``small'', ``medium'', ``large''}, respectively.
Then, we calculate the area of each combination of text wrapping $S_i$ and font size $F_j$.
By minimizing the waste of empty space, we calculate the ratio occupied by the text on the text block and determine the combination with the largest ratio as the optimal text typography.
Finally, we use optimized wrapping text to imply the text alignment ($S_i \Rightarrow A$), i.e. text is center-aligned if it is a single line, otherwise, left-aligned.
For example, in Fig.~\ref{fig:post}-B, the text is generated as the class of ``title'', so we highlight the text with \textit{``large''} font.
In Fig.~\ref{fig:post}-C, the ``subtitle'' text is post-processed to \textit{``small''} font.

\subsubsection{UI Guidelines}
UI design is not a simple recipe, it is subject to design rules and guidelines to match the psychology of human perception, that machine may not aware of.
To investigate the presence of design flaws in LLMs' generated UI wireframes, we conduct a small-scale study on randomly 100 generated UI wireframes.
Note that this study just aims to provide an initial analysis of possible UI design flaws in the generation of LLMs, and a more comprehensive study would be needed to deeply understand it.
Following the Card Sorting~\cite{spencer2009card} method, we summarise three potential flaws based on existing UI/UX design books and websites such as Design Pattern Gallery~\cite{neil2014mobile}.
To enhance the wireframes, we apply tailored heuristics that incorporate human design knowledge:

\textbf{a) Occlusion:} the textual information or element is occluded by other elements. This might be caused by the improper generation of the element’s width and height. To resolve this, we add a small margin between them without affecting other elements, as shown in Fig.~\ref{fig:post}-D.

\textbf{b) Duplication:} the elements are similar in many aspects of properties, such as size, class, position, text, etc. The possible reason could be that LLMs lack a very long memory~\cite{zhang2021commentary}. To resolve this, we compare the properties of each element and empirically set a threshold to determine if they are similar to be removed, for example, the redundant and overlapped buttons in Fig.~\ref{fig:post}-E.

\textbf{c) Out-of-bound:} the size of the element exceeds the bounds of the UI. It usually occurs with text elements that require relatively wide borders. To resolve this, we trim the size of the element, as shown in Fig.~\ref{fig:post}-F.

\section{Evaluation}
\label{sec:evaluation}
In this section, we conduct experiments to demonstrate the effectiveness of our fine-tuned LLMs in generating mid-fidelity UI wireframes.

\subsection{Research Questions}
\noindent\textbf{RQ1: (Performance of Fine-tuning) Does the optimization of fine-tuning correlate with better UI wireframe generations?} 
For RQ1, we evaluate the effectiveness of fine-tuning the LLMs, compared with widely-used in-context learning baseline approaches (e.g., zero-shot and few-shot).

\noindent\textbf{RQ2: (Performance of Variations) Do the variations of the LLMs result in better UI wireframes?} 
For RQ2, we present the comparison among the variations of GPT models (e.g., Curie, Babbage, Ada) to demonstrate the performance of using Turbo as the model. 

\subsection{Testing Data}
We collect the data as discussed in Section~\ref{sec:dataset} as our experimental dataset.
Since we leverage 1,000 sample data to fine-tune the LLMs, we first remove these data to avoid potential bias.
To evaluate the LLMs' generalizability and diversity, we randomly select 2 apps from each category.
In total, we collect 100 UI textual descriptions as the input prompts to generate UI wireframes.
Note that we do not use the corresponding UI screens as the ground-truth because the generative LLMs may create reasonable UI wireframes but deviate from the ground-truth.

\subsection{Annotation Methodology}
We recruit two people who had backgrounds in UI/UX design and art practice to rate the UI wireframe designs generated by our LLMs.
At the beginning of the experiment, we first give them an introduction to our study and ask them to read UI/UX design books for 30 minutes to deepen the design principles. 
Given the natural language descriptions and a grid of UI wireframe designs, annotators are then asked to independently rate which designs in the grid are either significantly better generations or significantly worse generations.
Note that we randomly shuffle the designs in the grid, so that the annotators do not know which design is generated from our model or baselines.
All annotators are compensated \$20/hour for however long it takes them to complete the task.

\begin{figure*}
	\centering
	\includegraphics[width = 0.7\textwidth]{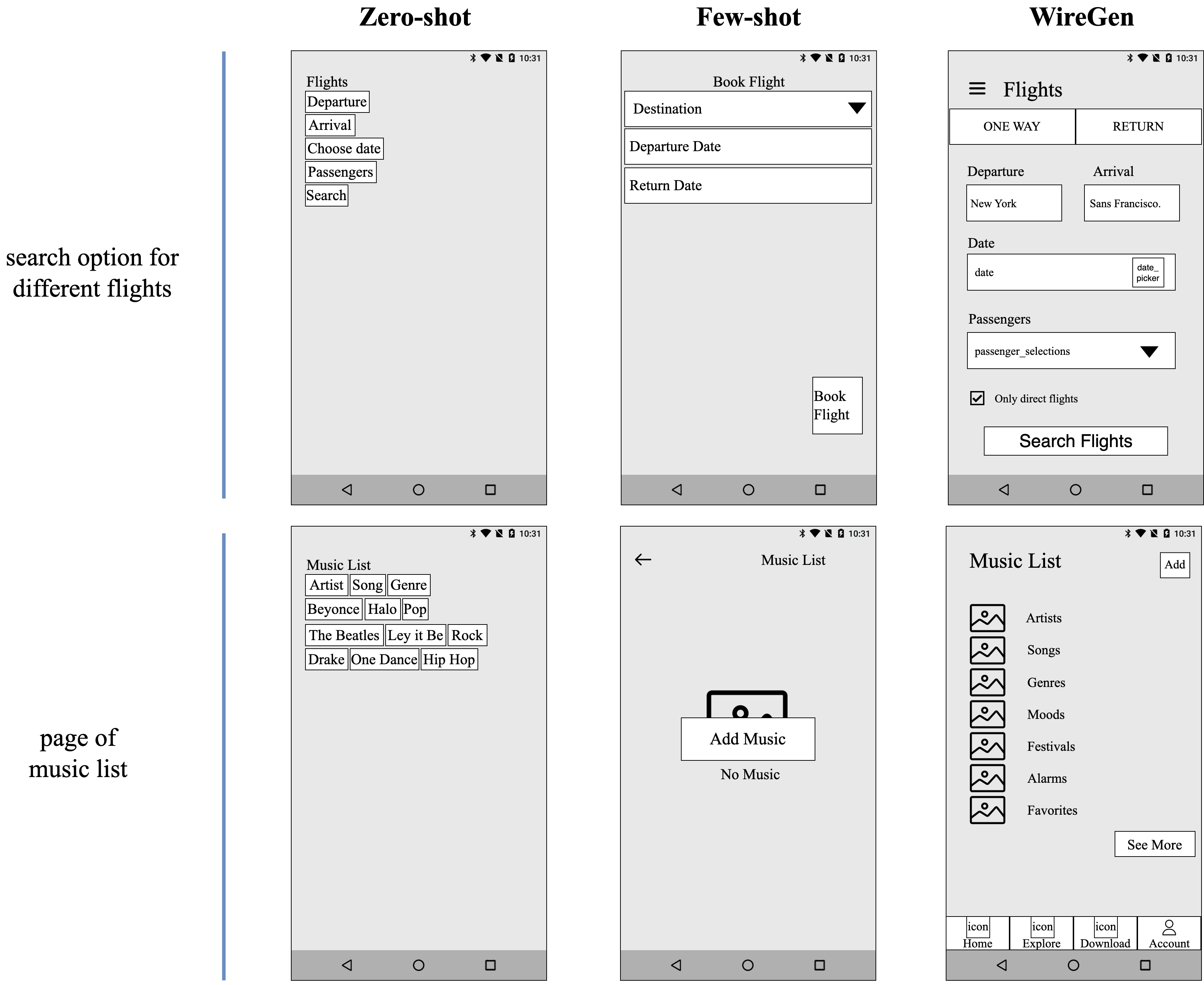}
	\caption{Examples of prompting UI wireframe generations between zero-shot, few-shot, and our fine-tuned LLMs.}
	\label{fig:result1}
\end{figure*}

\subsection{RQ1: Performance of Fine-tuning}
\subsubsection{Baselines}
To demonstrate the advantage of fine-tuning to master the domain-specific task of UI wireframe generation, we compare it with two widely-used in-context learning methods as baselines, including zero-shot learning and few-shot learning. 
Note that all the experiment settings (e.g., model, hyperparameter, etc.) are the same.

\textbf{Zero-shot learning:}
It predicts the results without any training samples.
The general idea behind zero-shot learning is the LLMs train on a wide collection of different databases and workloads and can thus generalize to a completely new task and workload without the need to be trained particularly on that task.

\textbf{Few-shot learning:}
It refers to giving a few demonstrations of the task as conditioning to allow the LLMs to predict the results of new tasks.
Typically, the demonstration has a prompt and a desired result (e.g., a login UI wireframe -> [UI wireframe]), and few-shot works by giving $K$ examples of prompt and result, and then one final prompt, with the LLMs expected to predict the result.
We set $K$ in the range of 1 to 2, as this is how many examples can fit in the LLMs’ maximum input tokens (4,096).

\subsubsection{Results}
From the annotations we collected, the generations from our \tool receive much better ratings than that of other baselines, e.g., 85.5\% of our generation is rated as significantly better on average.
In contrast, the generations receive an average of 51.5\% and 44.5\% significantly worse ratings for zero-shot and few-shot, respectively.
As the ratings come from the subjective nature of the annotators, we further check the agreement between them by calculating Cohen's kappa~\cite{mchugh2012interrater}.
We observe two annotators share an inter-rater reliability of 0.32, indicating a high agreement on scoring across generations.

Fig.~\ref{fig:result1} shows some UI wireframe generations from the baselines and our \tool.
We can see that zero-shot can generate some relevant elements due to its strong in-context learning ability.
For instance, it generates ``departure'', ``arrival'', ``passengers'' for a flight page and different music genres for a music page.
However, without any fine-tuning and example prompts, zero-shot does not consider the elements' properties (e.g., size, alt-text, etc.) or the overall design of the layout.
We can also observe that the few-shot is a double-blade.
Providing a small number of samples in the prompts can help the LLMs grasp the task better, leading to improved elements and layouts in the UI wireframes, which address the issues of zero-shot.
However, the limited number of samples may not be enough for the LLMs to generalize to the complex UI wireframing task, leading to some subpar layouts as shown in Fig.~\ref{fig:result1}.
Our \tool overcomes these limitations by the optimization of fine-tuning over thousands of UI samples, allowing the LLMs to inherit its in-context ability and master UI-specific understanding, resulting in better UI wireframe generations.

\begin{figure*}
	\centering
	\includegraphics[width = 0.85\textwidth]{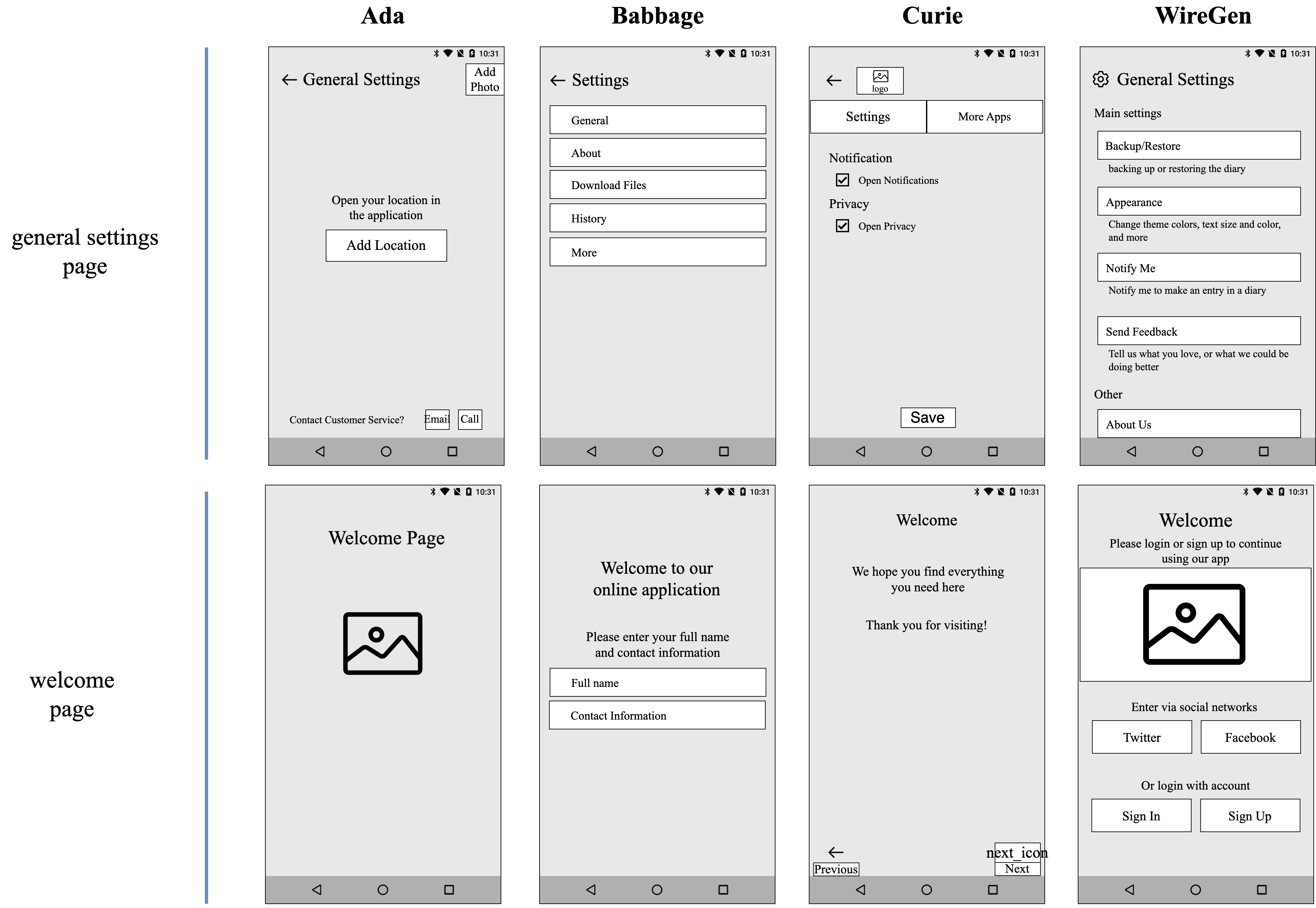}
	\caption{Examples of the comparison of different models, including Ada, Babbage, Curie, and our model \tool.}
	\label{fig:result2}
\end{figure*}

\subsection{RQ2: Performance of Variations}
\subsubsection{Baselines}
As our implementation of LLMs GPT offers different models to support different levels of tasks, we set up the other models, including Curie, Babbage, and Ada, as the baselines to compare with our model.
To guarantee the fairness of the experiment, we fine-tune these models on the training dataset using the same training settings as discussed in Section~\ref{sec:finetune}.

\subsubsection{Results}
As expected, our model Turbo performs more significantly better UI wireframes compared with other baseline models, e.g., 76.5\% vs 11.5\%, 10.5\%, 1.5\% for Curie, Babbage, and Ada, respectively.  
Both Curie and Babbage show the tendency to generate average UI wireframes (78\% and 82.5\%), while Ada underperforms with 81.5\% of UI wireframes rated to be significantly worse.
We report the inter-rater reliability score of 0.28 in Cohen's kappa, which represents a fair agreement, which we think is valid considering the highly subjective nature of the task (picking ``better'' or ``worse'' UI wireframe designs).

Fig.~\ref{fig:result2} shows some examples generated by our model (Turbo) and the baselines (Curie, Babbage, and Ada).
Our model, Turbo, outperforms the other models in terms of generating detailed and granular UI wireframes.
As seen in the examples, Ada generates the simplest and fewest elements, while Babbage and Curie generate some relevant information but are not as detailed as Turbo. 
For instance, our model is able to generate precise subtitles for each title, like ``Appearance'' with a subtitle of ``Change theme colors, text size and color, and more''.
This is unsurprising, given the fact that Turbo is the most powerful model, while the other baseline models are simplified versions that are trained on a smaller corpus of data with limited knowledge.
We believe the performance could be further boosted with the more advanced LLMs in the future.

\section{User Study}
\label{sec:user_study}
In language, there are many ways to say the same thing in different words.
We conduct a user study to gain insights into the usefulness of our model in assisting designers with UI wireframe generation with real-world descriptions.

\subsection{Participants}
We invite five UI/UX designers to participate in our experiments by word-of-mouth and through online advertisements on social media and design communities.
\begin{itemize}
    \item P1: Visual designer, 6 years of working experience, now working at a large multinational company, responsible for international app design and design innovation.
    \item P2: Interaction designer/product designer from a medium-sized company, with 4 years of working experience. Her work mainly focuses on smart devices.
    \item P3: Designer from a large IT company for 2 years with a focus on mobile advertising user experience design.
    \item P4: Visual designer from a well-known UI/UX design sharing community. He has 4 years of visual design experience.
    \item P5: Interaction designer/researcher from a medium-sized company, with 3 years of working experience. Her focus is on researching UI design intelligence.
\end{itemize}

\subsection{Procedure}
The user studies were conducted online via the Zoom platform.
Each participant took about one hour to complete. 
Each study session began with a short interview in which the participants were asked about their professional background and their experience in designing UI/UX examples.
We then gave them an introduction to our study and also an example to try.
After the introduction, we asked participants to use \tool in two tasks.
First, they were given a common wireframe design prompt (i.e., a login page) and were allowed to edit the description prompt to generate a practical mid-fidelity UI wireframe.
This task allowed us to observe how participants designed the UI wireframes and which of the descriptions they retained or modified.

After the first task, participants started the second task after a 5-minute break.
For the second task, we asked participants to freely use the tool, bringing in their own design scenarios.
Participants were allowed to use wireframe sharing tools (e.g., Moqups~\cite{web:moqups}, Dribbble~\cite{web:dribbble}, etc.) and intelligence tools (e.g., layout generation tools such as LayoutTransformer~\cite{gupta2021layouttransformer}, text-to-image generation tools such as DALLE~\cite{ramesh2022hierarchical}, Stable Diffusion~\cite{rombach2022high}, MidJourney~\cite{web:midj}, etc.).
We did not restrict how and when they used our tool for generations.
This task allowed us to observe how participants used our tool during the real-world UI wireframing process and a rough comparison to other tools.
In both tasks, we asked participants to share the screen of the interface.

Lastly, we conducted a short survey containing three questions asking participants to rate our tool \tool: 1) the effectiveness of the UI wireframe generations for inspiration; 2) the generated UI wireframes are related to the descriptions;  3) the diversity of the generations.
Each question was rated on a 5-point likert scale (1:strongly disagree and 5:strongly agree). 
At the end of the session, we conducted a short interview to collect open-ended feedback, such as how they used the \tool, how they modify the wireframe generated by the tool, how they would adopt \tool to their practice, and how they expected to improve.

\begin{figure*}
	\centering
	\includegraphics[width = 0.99\textwidth]{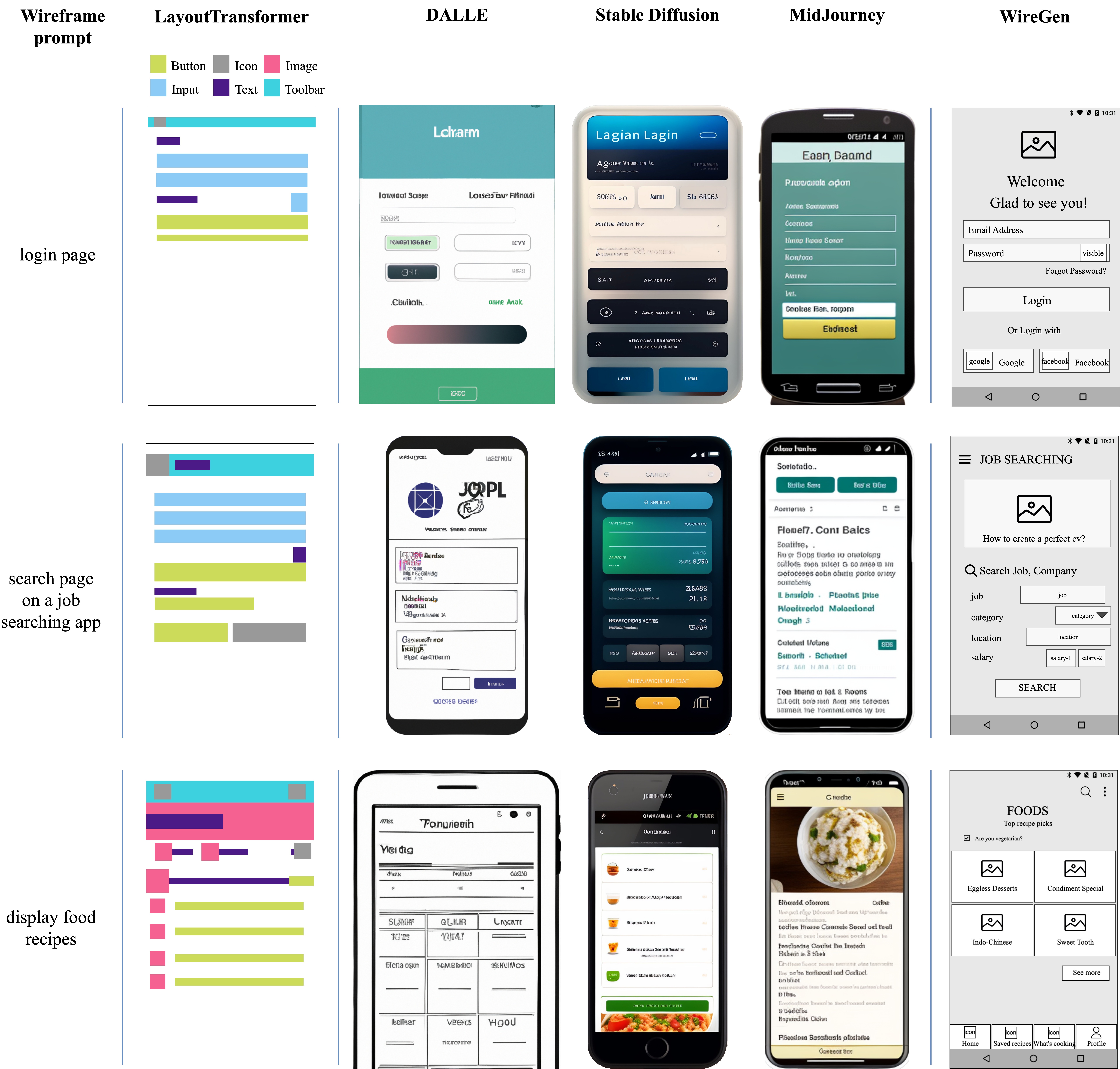}
	\caption{Examples of three design prompts in the user study. We compare the our generations with the layout generation model (LayoutTransformer) and three text-to-image models (DALLE, Stable Diffusion, MidJourney). The wireframes generated by LayoutTransformer are too abstract for understanding. Meanwhile, the wireframes generated by the text-to-image models often contained irrational text and design inconsistencies, which could hinder the process of gaining inspiration. In contrast, our model \tool generates better wireframes.}
	\label{fig:comparison}
\end{figure*}

\subsection{Results}
Overall, participants appreciated the usefulness of our \tool, which generates mid-fidelity wireframes by simply describing them in words to facilitate the UI design process.
In addition, the statistical analysis revealed a high level of satisfaction for all metrics.
We presented the results in detail below.

\subsubsection{Metric performance}
Most of the participants (4.4/5.0) agreed the wireframes can be effectively generated by our tool for inspiration, echoing the participants' behaviors in the first task to design a login page.
They edited and modified an average of 2.4 times to the prompt to get satisfactory UI wireframes in the first task.
P3 mentioned:
\textit{``Login pages usually have the general functionality of `Username' and `Password' fields, which is effectively generated without any prompt modification. Looking back at my previous design experience, an indispensable functionality of a login page is the third-party authentication service, such as Google. So, I added an additional description to the prompt (a login page with Google authentication). I found the generations very effective and interesting, that some designs use buttons with textual descriptions (e.g., `Sign in with Google'), and some use icon buttons, depending on the layout of the UI design.''}

Most of the participants (4.0/5.0) agreed the wireframes generated by \tool are relevant to the description prompts.
For instance, P1 mentioned:
\textit{``I can see some wireframe design patterns related to my description. For example, when I prompted the login page, the generated wireframe designs (mostly) consist of `login' and `register' buttons''.}
P2 described a particular use case of using our tool:
\textit{``I was working on creating a search page for a job searching app, so I prompted the description in the tool to gain inspiration (Fig.~\ref{fig:comparison}). I was amazed at how the UIs were generated. It closely mimics the real UI designs, such as a job, category, location, salary, etc.''}

Most of the participants (3.6/5.0) admit that our tool can provide diverse UI wireframes.
P3 mentioned the diverse UI element designs of using buttons and icons to support third-party authentications.
Two participants noted the diversity of the UI layout and P1 provided an example:
\textit{``In my last job of designing
a food app, I needed to decide on the display style of the recipes. I prompted the `display food recipes' in the tool (Fig.~\ref{fig:comparison}). Checking through the generations, I discovered two display styles, grid, and list, that match real UIs and human perception.''}
Besides, two participants appreciated the diversity in functionality.
P1 explained:
\textit{``I especially noticed the different functionalities between the wireframe generations over several prompts. It typically varies in menu functionality. On the one hand, I found the menu interaction in the top left corner is displayed as a three-line icon to save more design space. On the other hand, I found the interaction of the menu at the button navigation bar to show the functionality more explicitly.''}

\subsubsection{Comparison performance}
All participants agreed that \tool has significant advantages over other tools.
First, the existing wireframe sharing tools are not able to tailor designs to the needs, while \tool allows for the addition of descriptive prompts to generate more specific UI wireframes.
For example, P2 describes a practical use case when he wants to add a message functionality to the job searching page:
\textit{``When I searched `job searching' or `message function' in Dribbble, there are many different kinds of returning results. I had to manually filter out irrelevant ones, which takes time. But when I zoomed in on a more specific query like `a job searching UI with message functionality', no customized results were retrieved. In contrast, with \tool, it solves my need to generate more customized designs and I could add more descriptions to flesh out the designs.''}

Second, while there are many tools aimed at helping designers gain inspiration on UI wireframes, they are either low-fidelity (such as layout generation) or impractical (such as text-to-image generation).
As P5 stated:
\textit{``I attempted to use the recent state-of-the-art image generation techniques, such as DALLE, Stable Diffusion, and MidJourney, to create UI wireframe designs. The results were not useful, due to 1) unreasonable text and 2) design incoherence like distorted shapes, as shown in Fig.~\ref{fig:comparison}.''}
P5 also highlighted the limitations of layout generation tools, explaining that:
\textit{``The generated layouts are too abstract to understand in Fig.~\ref{fig:comparison}. The designers still have to do a lot of work mentally to imagine the contents in their heads. And the actual contents may impact the overall design. For example, a layout originally intended for image and text side-by-side may change to a top-bottom layout if the actual image is too large or the text too long.''}
In comparison, the generated wireframes by \tool were perceived as a gateway to gather ideas of actual details of the content, presented in the context of the designer’s concept.

\subsubsection{Area to improve}
The participants responded positively to \tool and gave several directions to improve our tool.
First, they wondered if we could develop our tool as a plugin to integrate with popular wireframe sketching software, such as Adobe PhotoShop~\cite{web:photoshop}, Sketch~\cite{web:sketch}, so that they could modify the wireframe and serve with more powerful editing capabilities.
Second, they suggested the addition of representative images in the wireframes, rather than relying solely on alt-text to convey the image content. 
This could lead to the creation of a high-fidelity wireframe.
We believe that it is not difficult to retrieve the representative images once we have a large image resource database.
In the future, we will continue to improve our tool for better performance in generating better wireframes.

\section{Discussion}
In Section~\ref{sec:evaluation} and Section~\ref{sec:user_study}, we evaluate and examine \tool’s effectiveness and usefulness for automatically generating mid-fidelity UI wireframes from natural language descriptions.
We see several opportunities to improve the performance of our LLMs. 
For example, we see that training on a large dataset can deliver better results.
However, due to budget constraints, we only use 1,000 data for training.
Once we have enough budget, we can improve the performance of the LLMs by incorporating more data.

The LLMs are obviously not omnipotent, and still, fail to provide ready-to-use outputs in many cases.
To enhance the raw wireframe generations, we implement several post-processing methods (Section~\ref{sec:post}) based on a small-scale study, including incorporating semantic icons, adjusting text typography, and following UI guidelines.
However, this study was limited in scope and merely aimed to provide a basic analysis of the potential UI design flaws produced by GPT-3. Further research is needed to fully understand the design limitations of LLMs and to develop more effective solutions.

Our work focuses on using UI screens and their corresponding view hierarchy information to construct UI-specific prompts.
However, UI screens have various other modalities, including app information, interaction context, etc, which are left unused in our study.
In future work, we aim to enhance the prompts by incorporating representations of multiple modalities, thereby improving performance.

Our work focuses on generating UI wireframes from a single description.
However, UI design often involves multiple iterations and accumulations of ideas.
To streamline this process, we aim to integrate chat interactions in the future.
This could be achieved through inspired by ChatGPT~\cite{web:chatgpt}.
For example, a designer could interact with the chatbot to design a login page with Google authentication service: ``[Designer]: I want a UI wireframe design of a login page.''; [Bot]: <UI Wireframe>; [Designer]: I want to add a Google authentication."; [Bot]: <UI Wireframe+Google>", simulating a human-like design process through dialogue.
In the future, we aim to enhance the interactivity of our design process, making it more conversational in nature.

A straightforward extension of our work would be the retrieval of UI design examples.
Numerous studies~\cite{herring2009getting,huang2019swire,bunian2021vins} have demonstrated the usefulness of low-fidelity wireframes to retrieve existing UI design examples from datasets, thereby facilitating UI design.
Our mid-fidelity wireframes could further enhance this retrieval process by incorporating semantic information such as text and icon semantics.
To achieve this, we could employ the mature Screen2Vec method~\cite{li2021screen2vec} to encode visual and textual information into an embedding vector.
By comparing the embedding vectors between the generated UI wireframe and the designs in existing UI datasets, we could retrieve semantically similar UI designs.
This would help designers broaden their perspectives and gain inspiration.
Future work could leverage our work into these potential applications to reduce the time and resources required for UI design, and ultimately improve the user experience.
\section{Conclusion}
This paper introduces \tool, a new method for exploring human-AI collaboration in mid-fidelity UI wireframe creation.
Inspired by the success of generative Large Language Models in creating innovative ideas, we purpose to inherit LLMs' in-context ability from billions of web resources to create mid-fidelity UI wireframes.
Since LLMs are not designed for UI understanding and creation, we fine-tune it with thousands of UI data, including screens and view hierarchies, to help the LLMs master UI design knowledge.
With the fine-tuned LLMs, we prompt a simple description of design intent to generate a raw wireframe and apply several post-processing methods to make it more intuitive, engaging, and interactive.
We set up experiments to demonstrate the performance of \tool in generating mid-fidelity UI wireframes, significantly outperforming the widely-used baselines and ablated models.
Additionally, we carry out a user study to confirm the practical usefulness of \tool, demonstrating the potential to facilitate the process of UI design.

\bibliographystyle{ACM-Reference-Format}
\bibliography{reference}

\end{document}